# Surfactant-triggered disassembly of electrostatic complexes probed at optical and quartz crystal microbalance length scales

N. Schonbeck[a]*, K. Kvale[a], T. Demarcy[a], J. Giermanska[b],
J.-P. Chapel[b]* and J.-F. Berret[a]*

[a]Matière et Systèmes Complexes, UMR 7057 CNRS Université Denis Diderot Paris-VII, Bâtiment Condorcet
10 rue Alice Domon et Léonie Duquet, F-75205 Paris, France
[b]Centre de Recherche Paul Pascal (CRPP), UPR CNRS 8641, Université Bordeaux 1, 33600 Pessac, France

A critical advantage of electrostatic assemblies over covalent and crystalline bound materials is that associated structures can be disassembled into their original constituents. Nanoscale devices designed for the controlled release of functional molecules already exploit this property. To bring some insight into the mechanisms of disassembly and release, we study the disruption of molecular electrostatics based interactions via competitive binding with ionic surfactants. To this aim free-standing micron-size wires were synthesized using oppositely charged poly(diallyldimethylammonium chloride) and poly(acrylic acid) coated iron oxide nanoparticles. The disassembly is induced by the addition of sodium dodecyl sulfates that complex preferentially the positive polymers. The process is investigated at two different length scales: the length scale of the particles (10 nm) through the Quartz Crystal Microbalance technique, and that of the wires (> 1 µm) via optical microscopy. Upon surfactant addition, the disassembly is initiated at the surface of the wires by the release of nanoparticles and by the swelling of the structure. In a second step, erosion involving larger pieces takes over and culminates in the complete dissolution of the wires, confirming the hypothesis of a surface-type swelling and erosion process.

# Introduction

Electrostatic assembly describes the process by which oppositely charged species associate via the pairing of electrostatic charges. This approach has attracted much attention during the last years because it is non-specific and can generate a wide variety of structures.[1] Among these structures, thin films of polyelectrolyte multilayers[2,3] and polyionic complex micelles[4,5,6] are prominent examples. These assemblies are used in multiple applications in materials science and nanotechnology, including coatings, packaging and drug delivery systems.

An important feature of electrostatic assembly is that the building blocks (e.g. polymers, particles, proteins, biological molecules) are associated through non-covalent bounds. The components are held together by multiple electrostatic interactions, and it is possible to unbind them through suitable strategies. Activated at specific sites, the disassembly can be used to control the release of drugs and of functional molecules.[7, 8, 9] Different approaches to disassemble polyelectrolyte multilayers have been reported.[10, 11] They involve *i)* the change of physico-chemical environment such as salt or pH,[12, 13, 14] *ii)* the disruption of internal molecular interactions via competitive binding,[14, 15, 16, 17, 18, 19, 20] *iii)* electrochemical methods[21] and *iv)* the chemical reduction of active bonds such as disulfide links.[7, 22]

In this work, we focus on the second process listed above. Competitive binding is a phenomenon in which an ion-containing structure is destabilized through the addition of a third (charged) species. It is assumed that this additional species complexes preferentially one of the two polymers forming the multilayers, and that the initial matrix dissolves via the removal of one of its component. So far, few studies of applicative interest have been conducted to explore this phenomenon.[20] Some experiments have been performed using charged proteins,[15, 23, 24, 25] surfactants[14, 16, 17, 18, 26] and multivalent counterions.[19] Ramos *et al.*[18] made use of cationic alkyl ammonium bromide surfactants to remove predetermined patches from a poly(diallyl dimethyl ammonium chloride)/ poly(sodium 4-styrene sulfonate) (PDADMAC/PSS) Layer-*by*-Layer (L*b*L) structure, and to obtain patterned substrates for immobilization applications. Interestingly, these authors found that not all polyelectrolyte pairs reacted with the surfactant studied. More recently, Kang and Dähne investigated the response of polyelectrolyte microcapsules to cationic surfactants and found a strong dependence on the polyelectrolyte pair.[16] For capsules composed of 4 double layers of PDADMAC/PSS, the fluorescently labeled objects disintegrated completely upon addition of dodecyl trimethylammonium bromide (DTAB). In Kang and Dähne's experiments, the kinetics of disassembly was monitored by the measure of the fluorescence decay. Disassembly times were found to vary widely from a few seconds to hundred of seconds, depending on the surfactants and polyelectrolytes used. Competitive binding was also studied using ferrocyanide multivalent ions to dissolve the multilayer on which living cells were grown.[19] This approach of cell sheet engineering has direct applications in regenerative medicine. Concerning the microscopic process of multilayer disassembly, several mechanisms were proposed, but a final description is still missing.[10] *Does the erosion proceed throughout the bulk of the multilayer, or does it start from the surface and progressively disrupt the film?* Moreover, the approaches described above were mainly conducted on supported thin films and free-standing capsules made through the sequential adsorption of oppositely charged polyelectrolytes. The growth of such L*b*L films is generally governed by the intrinsic charge compensation occurring at each step and giving rise to a (bulk) thin film globally neutral with an outermost surface region where counterions balance excess polymer charges.[27, 28, 29] The generated thin films are hence close to charge stoichiometry.

In the present study, we put under scrutiny micron-size wires obtained by electrostatic co-assembly using oppositely charged particles and polymers. This bottom-up technique produces wires with diameters between 0.1 – 1 µm and lengths between 1 and 200 µm.[30, 31, 32] The disassembly of electrostatic complexes was investigated using this particular type of objects for the following reasons: *i)* wires are indeed highly anisotropic, and any variation of their length is easily detected. *ii)* Wires are also in the micron range and have excellent contrast in optical microscopy. *iii)* Wires are freestanding objects and structural changes will not be perturbed by the interactions with the substrate. Taking benefit from such characteristics, we report and monitor here the swelling and erosion of electrostatic wires upon addition of ionic surfactants at two different length scales: at the scale of the wires (> 1 µm) via optical microscopy and at the scale of the particles (> 10 nm) through the Quartz Crystal Microbalance (QCM) technique. To evaluate the role of electrostatic charges, positively and negatively charged wires with similar dimensions were prepared. We propose a complete scenario of the disassembly process.

# Experimental

## Synthesis of magnetic wires by electrostatic assembly

Wires were formed by electrostatic complexation between oppositely charged nanoparticles and polymers.[30, 31, 32] The particles were 8.3 nm iron oxide nanocrystals (γ-$Fe_2O_3$, maghemite) synthesized by polycondensation of metallic salts in alkaline aqueous media.[33, 34] The particles were characterized by various techniques including vibrating sample magnetometry, dynamic light scattering and Transmission Electron Microscopy (TEM) (see also Supplementary Information S1-S2). The magnetic diameter determined by magnetometry (8.3 nm) was slightly lower than that obtained by TEM (9.3 nm). This difference was attributed to the existence of a disordered oxide layer at the outer surface of the particles. Both techniques resulted in obtaining lognormal distributions of the particle size with polydispersity 0.20 [34]. The polydispersity of particles is defined here as the ratio between the standard deviation of the distribution and the average diameter. The molecular weight of the nanocrystals was measured using static light scattering at $M_W = 5.8 \times 10^6$ g mol$^{-1}$. To improve their colloidal stability, the cationic particles were coated with $M_W = 2100$ g mol$^{-1}$ poly(sodium acrylate) ($PAA_{2K}$, Aldrich) using the precipitation-redispersion process.[34, 35, 36] This process resulted in the adsorption of a highly resilient 3 nm polymer layer surrounding the γ-$Fe_2O_3$ particles. The hydrodynamic diameter of the coated particles was determined by dynamic light scattering and found to be 30 nm (value of the second cumulant). Acido-basic titration experiments performed on the $PAA_{2K}$ coated particles allowed for the number of adsorbed polymers per particle and the charge density to be evaluated. For the 8.3 nm particles, these values were 470 ± 30 chains and 25 ± 3 electric charges per square nanometer.[37]

The co-assembly process used to create the magnetic wires followed a bottom-up approach, where the building blocks were the iron oxide nanoparticles and the "gluing" agent a highly charged polycation, poly(diallyldimethylammonium chloride) (PDADMAC, Aldrich) of molecular weight $M_W < 100000$ g mol$^{-1}$.[32] Fig. S3 (Supporting Information S3) illustrates the desalting protocol designed for the synthesis of the magnetic wires. In the polymer/particle mixture used for the dialysis, the relative amount of each component was monitored by the charge ratio Z. Z was defined as the ratio between the anionic charges borne by the particles and the cationic charges carried by the polymers. According to this definition, $Z = 1$ describes the isoelectric solution. In the current study, we compare the behavior of cationic and anionic wires and their interactions with ionic surfactants. Wires were prepared apart from the isoelectric charge at $Z = 0.3$ (excess of polymers) and $Z = 2.5$ (excess of particles). The shelf life of the co-assembled structures was of the order of several years. In the present experiments, the wires were found to have an average length of 20 µ$m$ and a polydispersity of 0.6. Lengths were typically comprised between 1 and 50 µm.

Surfactant and Control

Experiments made use of two different ionic surfactants: anionic sodium dodecyl sulfate (SDS) and cationic dodecyltrimethylammonium bromide (DTAB). Surfactant solutions were prepared at weight concentrations of c = 0.2, 1, 5, and 30 wt. % for SDS, and at c = 10 wt. % for DTAB. The critical micelle concentration (CMC) for the surfactants was 8.2 mM (0.24 wt. %) for SDS and 15 mM (0.46 wt. %) for DTAB in pure water at 25 °C. Above the CMC and at the concentrations investigated, SDS and DTAB molecules self-assemble into 4 nm spherical micelles.[38] Finally, a 2 M solution at a pH of 8 of an inorganic crystalline salt, ammonium chloride ($NH_4Cl$, Aldrich), was used as a control.

Optical Microscopy

A POCmini-2 Chamber System® (PeCon, Erbach, Germany) connected to an inverted optical microscope was adapted to visualize the disassembly process at the scale of the wires (Fig. 1).

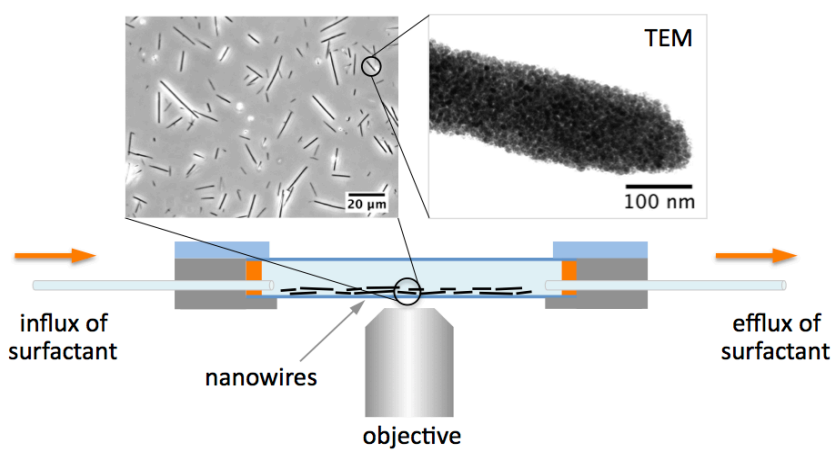

*Figure 1*: *Schematic representation of the close perfusion system (PeCon, Erbach, Germany) used for the monitoring of wire disassembly. Surfactant or salt solutions are injected through a pipe on the left-hand side via plastic tubing and, after interacting with the wires collected on the right hand-side in a beaker. Wires were exposed to a steady concentration as a function of time. Upper left panel: phase contrast microscopy of 20 µm wires adsorbed on a glass substrate. Upper right panel: extremity of a wire observed using Transmission Electron Microscopy (TEM). The small dark dots are 8.3 nm iron oxide particles.*

The perfusion system comprised two glass plates separated by an adapter with two 0.2 mm input/output pipes. The adapter was connected to a syringe pump (KD Scientific Infusion Pump) on one side (influx) and to a beaker on the other (efflux), and it was placed on the moving stage of the microscope. To ensure that the wires were exposed to a constant surfactant or salt concentration as a function of time, control experiments were carried out with a 0.6 mM methylene blue solution. It could be shown that for flow rates larger than 10 µl min$^{-1}$, the incoming solution exhibits an abrupt concentration front that moved from one side of the chamber to the other, and that molecular diffusion and concentration gradients were negligible. The wires adsorbed at the bottom of the chamber were hence subjected to a constant surfactant or salt concentration during the entire duration of the run. In this study, the experiments were performed between 10 and 1000 µl min$^{-1}$. In this range, similar time responses for the swelling and disassembly of the wires were obtained. Additional testing using an open perfusion system in which the injected solution was introduced at a pinpoint position provided results in agreement with the closed system, revealing again that the present device allowed retrieving the kinetics of disassembly.

All measurements with the controlled perfusion experiments were conducted on an Olympus IX71® inverted microscope connected to a QImaging EXi Blue® digital camera. The phase contrast images generated were acquired using MetaVue™ Research Imaging software at magnification intensities of 4×, 20× or 60× and assembled into stacks. Images were then linked together in sequence to depict a live-action response of the wires in the form of a film and finally processed with ImageJ software (http://rsbweb.nih.gov/ij/). The length of wires $L(t)$ was measured as a function of the elapsed time for at least 5 wires per experiment. In all movies recorded, $L(t)$ was found to increase rapidly and reach a steady state length, noted $L_f(t)$. At much longer times > 30 mn, depending on the experimental conditions, the wires eventually disassembled. Initial and final lengths $L_0$ and $L_f$ respectively were also measured for a minimum of 30 additional wires per experiment. In this study, the data are discussed in terms of relative length increase as a function of the time, $\Delta L_{Rel}(t) = (L(t) - L_0)/L_0$ and in the steady state, $\Delta L_{Rel} = (L_f - L_0)/L_0$. All the experiments were performed in triplicate and show identical responses.

Quartz Crystal Microbalance

A QCM set up with dissipation, QCM-D (*Q-Sense E4 instrument, AB, Sweden*) with silica-coated quartz crystals was used to monitor the disassembly process at the scale of the building blocks. In a typical experiment the crystal is excited at its fundamental resonance frequency ($f_0$) through a driving voltage applied across the gold electrodes. Any material adsorbing (resp. desorbing) onto the crystal surface induces a decrease (resp. increase) of the resonance frequency $\Delta f = f - f_0$. $\Delta f$ *is* directly related to the adsorbed mass per unit area (mg m$^{-2}$) through the Sauerbrey equation $\Delta m = -C \Delta f/n$, where $C$ is the Sauerbrey constant (0.177 mg s m$^{-2}$ for a 5 MHz quartz sensor) and $n$ is the overtone number. An indication of frictional losses due to viscoelastic properties of the adsorbed layer is provided by changes in dissipation $\Delta D = E_D/2\pi E_S$, where $E_D$ is the energy stored in the sensor crystal and $E_S$ is the energy dissipated by the viscous nature of the surrounding medium.[39] Positive and negative wires were adsorbed onto the silica crystal (~ 1 cm) outside of the QCM to avoid clogging the tubing system with such macroscopic objects. The crystal was then dried out and re-introduced into the cell insuring then a strong adhesion between the discrete wires and the crystal surface. A surface coverage around 5% was obtained as in the case of the optical experiment. The cell was then flushed with pure water to remove bubbles and let to equilibrate at 20 °C. Injection of SDS solution was then performed at a flow rate of 0.1 ml min$^{-1}$ for 5 mn in order to fill up the cell with surfactants at a given concentration (here twice the cmc). The flow was then stopped and $\Delta f$ and $\Delta D$ monitored as a function of the time.

# Results and Discussion
## Optical microscopy
*Short time scale*

Fig. 2 illustrates the swelling kinetics of wires treated with a sodium dodecyl sulfate solution. The experiment was performed using positively charged wires made from PDADMAC and from PAA$_{2K}$ coated iron oxide particles. The wires were loosely adsorbed at the lower side of the closed perfusion system and were submitted to the flow of a SDS solution at c = 0.96 wt. % at 100 μL min$^{-1}$. Phase-contrast microscopy snapshots (20×) at times t = 0, 20, 40, 140 and 500 s are displayed in the upper part of Fig. 2. The snapshots show that the swelling of the wires is rapid and occurs within the first 2 minutes. The movie of the entire sequence is available in Supporting Information (S4). The swelling is also accompanied by an increase of the flexibility of the wire. Submitted to the surfactant influx, the wires first become tortuous, and in some cases fold into flexible structures. In the study with SDS, the swelling and the onset of flexibility were found to be concomitant. Fig. 2f displays the relative length increase $\Delta L_{Rel}(t)$ of a $L_0$ = 80 μm wire (indicated by an arrow in Fig. 2a and 2e) as a function of the time. The temporal variation exhibits a strong initial increase, followed by a saturation plateau. The initial stage takes 40 s, and corresponds to a growth velocity of 2 μm s$^{-1}$. A close observation of the swelling using a 60× objective shows that the modification of the nanostructure is three-dimensional, *i.e.* it occurs in proportion similarly along the diameter and along the length. In the steady state ($t > 400\ s$), the length of the wire reaches the value $L_f = 150\ \mu m$, indicating a 94% increase. An analysis in terms of an exponential function is performed using:

$$\frac{L(t)}{L_0} - 1 = \Delta L_{Rel}\left(1 - exp\left(-\frac{t}{\tau_S}\right)\right) \quad (1)$$

where $\tau_S$ (= 30 s) is the characteristic time of the growth. Eq. 1 is found to be valid over a broad range of concentrations (Table I). In this stationary state, the wire appeared also less contrasted than in the initial state, indicating a decrease of the nanoparticle density inside the aggregates.

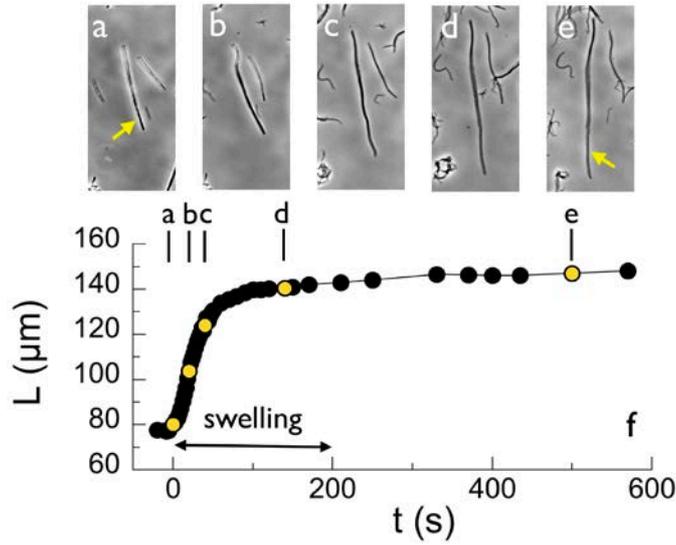

*Figure 2: Swelling kinetics of positively charged wires exposed to a sodium dodecyl sulfate solution at a concentration of 0.96 wt. % and a flow rate of 100 µL/min. a-e) Snapshots of wires at t = 0, 20, 40, 140 and 500 s, respectively. f) Relative length increase $\Delta L_{Rel}(t)$ of the wire shown by an arrow in a) and e). The length goes from 80 µm to 150 µm in the stationary state, corresponding to $\Delta L_{Rel} = 0.94$.*

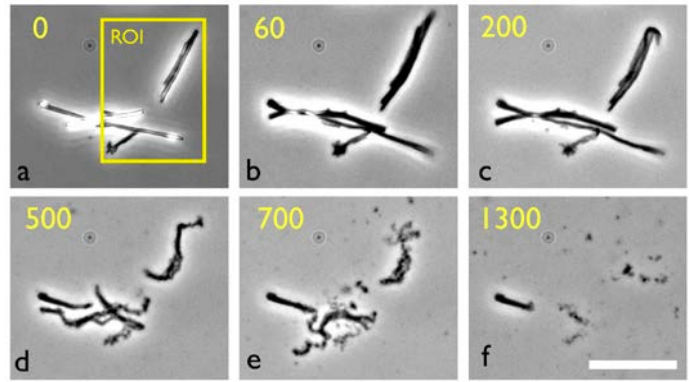

*Figure 3: Behavior of a cluster of wires loosely adsorbed on the glass substrate and submitted to similar conditions as in Fig. 2. After 500 s, the objects become flexible, break in several pieces and dissolve in the surrounding fluid. The bar is 20 µm. The rectangular Range Of Interest (ROI) indicated in a) outlines the field of view over which the pixel intensity is estimated to derive the proportion of wires present at time t. Note that a small portion of a wire strongly adsorbs and remains intact.*

*Long time scale*

To monitor the long-term behavior of the wires, a second experiment was designed where microscopy images are taken every 10 s, instead of every second as previously. Apart from the time-lapse, the experimental conditions were identical. Fig. 3a-f illustrates the behavior of a group of wires over a period of 1300 s. In these images, the interactions with SDS molecules lead again to a swelling and an increase in flexibility of the wires. A steep initial increase of the length is followed by a saturation around t = 200 s (Fig. 3c). Later, the structures become less dense and progressively disassemble, releasing in the surrounding aggregates and small filaments. To quantify the kinetics of the disassembly and determine the characteristic time of the process, a range of interest is outlined (indicated as ROI in Fig. 3a) and the pixel intensity in this ROI is determined as a function of the time. The signal is then translated into a quantity $P(t)$ that corresponds to the proportion of the wire in the field of view. Fig. 4 shows the proportion of a wire in the field of view as a function of the time, whereas the inset illustrates a swelling similar to that of Fig. 2. At first, $P(t)$ remains constant and then decreases to 0. This decrease is due to several effects, such as the reduction of the contrast, the increase in the flexibility and the slow disassembly of the structures. The decrease was adjusted using a stretched exponential of the form:

$$P(t) = \exp\left(-\left(\frac{t}{\tau_D}\right)^\alpha\right) \qquad (2)$$

where $\tau_D$ is the characteristic time for the disassembly, and $\alpha$ the stretching exponent. The continuous curve in Fig. 4 is obtained using $\tau_D = 800\ s$ and $\alpha = 2$. The experiment was repeated to evaluate the variability on the disassembly time. $\tau_D$ was found to range from 600 to 1800 s, depending on the diameter of the wires and on the degree of adsorption on the substrate (Table I).

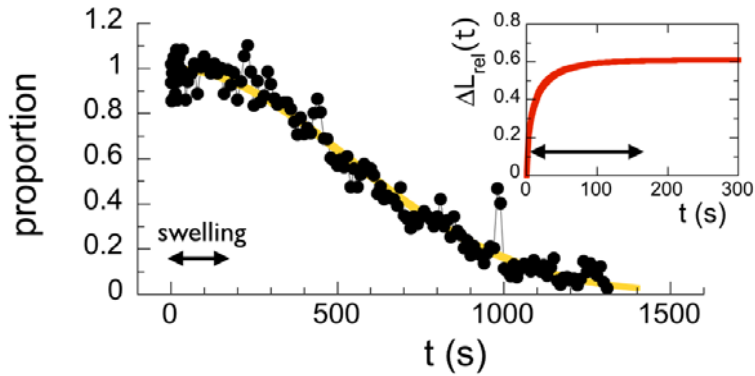

*Figure 4: Proportion of wires $P(t)$ in the range of interest defined in Fig. 3 as a function of the time. $P(t)$ varies between 1 when the wires are present and 0 when they are not. The proportion is calculated using a grey value threshold method. It is verified that the result does not depend on the choice of the ROI, nor on the threshold. The continuous curve results from least-square calculations using Eq. 2 and a disassembly time $\tau_D = 800\ s$. Inset: Relative length increase $\Delta L_{Rel}(t)$ of the wires shown in the ROI.*

Beyond the kinetics, optical microscopy performed in phase contrast and at high magnification (60×) also provides an accurate description of the erosion process at the length scale of the wires (> 1µm). Fig. 5 displays close-up views of a freestanding wire submitted to a dilute SDS solution (c = 0.96 wt. %). The same portion of wire is shown at different times, with image acquisition beginning 200 s after the onset of swelling. Initially, the wire exhibits some flexibility and good contrast, indicating a dense structure (Fig. 5a). Over time, the diameter increases and large pieces detach from the structure (arrows in Fig. 5c and 5d). The wire eventually breaks into several pieces, which dissolve in the surrounding. Fig. 5 suggests that the erosion is not a bulk-type process, but it is initiated at the surface, confirming the hypothesis of a surface-type swelling and erosion.[10] The snapshots in Fig. 5 also indicate that the erosion process is the major step in the disassembly and that at this stage it occurs through the release of micron-size aggregates. These large-scale aggregates were found to have various morphologies, from spherical to elongated, with sizes ranging between 0.5 and 2 μm. The erosion proceeds up to the complete de-structuration of the wire.

| Surfactant and salt | conc. | $\Delta L_{Rel}$ | $v_S$ (μm s$^{-1}$) | $\tau_S$ (s) | $\tau_D$ (s) |
|---|---|---|---|---|---|
| SDS | 0.96 wt.% | 0.81 | 1.4 | 30 | 600 – 1800 |
|  | 30 wt.% | 0.54 | 3.5 | 20 | 600 – 1800 |
| DTAB | 10 wt.% | 0.03 | 0.1 | 10 | ∞ |
| NH$_4$Cl | 2 M | 0.26 | 0.05 | 150 | ∞ |

*Table I*: *Results for the swelling and disassembly of positively charged wires. The wires were assembled via electrostatics using PDADMAC cationic polymers and PAA$_{2K}$-coated iron oxide nanoparticles. $v_S$ denotes the initial elongational velocity and $\tau_S$ and $\tau_D$ are the characteristic times for swelling (Eq. 1) and disassembly (Eq. 2) processes. For the profiles with SDS 30 wt. %, $\tau_S$ was determined from the $\Delta L_{Rel}(t)$ initial slope.*

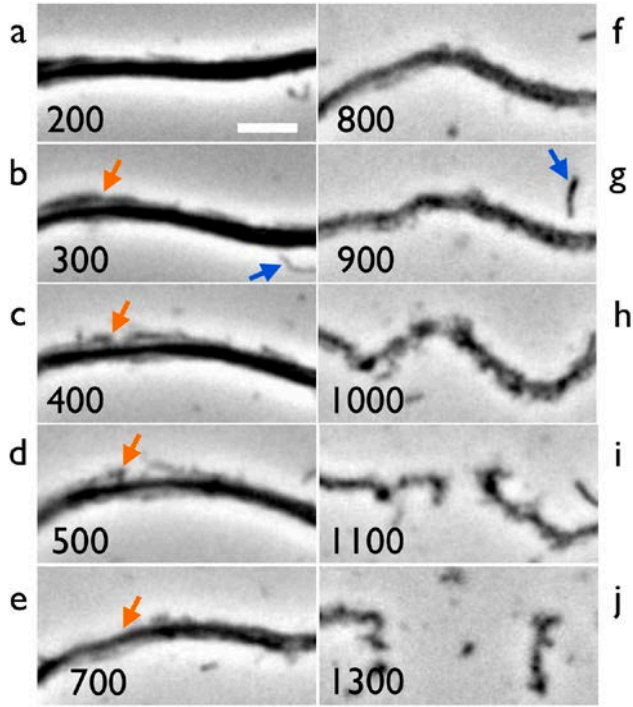

*Figure 5*: *Close-up views of a wire during disassembly. The bar in the first image is 3 μm. Note that as time evolves the outer surface becomes irregular, the diameter increases and large pieces detach from the structure. Detached aggregates indicated by arrows are about one micron in size.*

## Quartz Crystal Microbalance

To discriminate changes occurring at the submicron level, *i.e.* at the early stage of the swelling or inside the loose aggregates described previously, the Quartz Crystal Microbalance technique was used. Figs. 6a shows the frequency shift $\Delta f(t)$ for both negative and positive wires when exposed to a 0.48 wt. % SDS solution. A close inspection of the data shortly after the injection of SDS (inset in Fig. 6a) reveals a small but measurable negative undershoot, which is attributed to the adsorption of SDS unimers or micelles onto the wire structures. At this stage, SDS is in competitive binding with the poly(acrylic) coated iron oxide. The surfactants then replace some of the physical cross-links created during the wire synthesis, releasing then the nanoparticles from the network.

Very rapidly then, $\Delta f(t)$ turns positive, indicating a mass loss with a kinetics similar to that of optical microscopy. An analysis of the $\Delta f(t)$-increase was performed using:

$$\frac{\Delta f(t)}{n} = \Delta f_0 \left(1 - exp\left(-\frac{t}{\tau_{QCM}}\right)\right)$$

where $\Delta f_0$ is the steady state frequency and $\tau_{QCM}$ the characteristic time of the increase. It is found that $\tau_{QCM} = 19\,s$ for both types of wires, in good agreement with the swelling time $\tau_S$ from microscopy. The similarities in both transient behaviors indicate that the swelling is accompanied by the release of materials. It also suggests that erosion involves single iron oxide particles or at least aggregates smaller than the resolution of the optical microscope (here 500 nm). Optical images of the quartz sensor seeded with wires at different time frames also demonstrate that the disassembly of the wires occurs within the QCM cell (Fig. S7). As can be seen, the majority of the wires present disappear upon SDS addition. Some wires are however partially eroded and some remain intact, a result that could be inferred to the sample preparation of the QCM sensor. Furthermore, because negative wires are made with an excess of iron oxide as compared to positive ones, it is plausible that more nanoparticles are released, explaining the enhanced $\Delta f(t)$ signal found for the negative wires (Fig. 6a). The release of nanoparticles bearing multiple interacting sites will consequently loosen the wire network and trigger its swelling by water.

Fig. 6b displays the time evolution of the dissipation $\Delta D$ associated with the behavior of Fig. 6a. There, a noticeable increase of the dissipation followed by a weak and longstanding decrease is observed for positive wires, whereas a quite strong drop is found for negative wires. The transformation of a rather rigid structure into a looser and swollen network comes with an increase of the viscoelasticity of the structure. The softening of the structure agrees well with the initial dissipation growth. Furthermore, positive wires composed with an excess of elastic macromolecular chains (PDADMAC) will give rise to a more durable viscoelastic network upon release of the (few) nanoparticles, whereas negative wires, predominantly composed of rigid particles, will disassemble more rapidly upon nanoparticles release by a rapid drop of the dissipation.

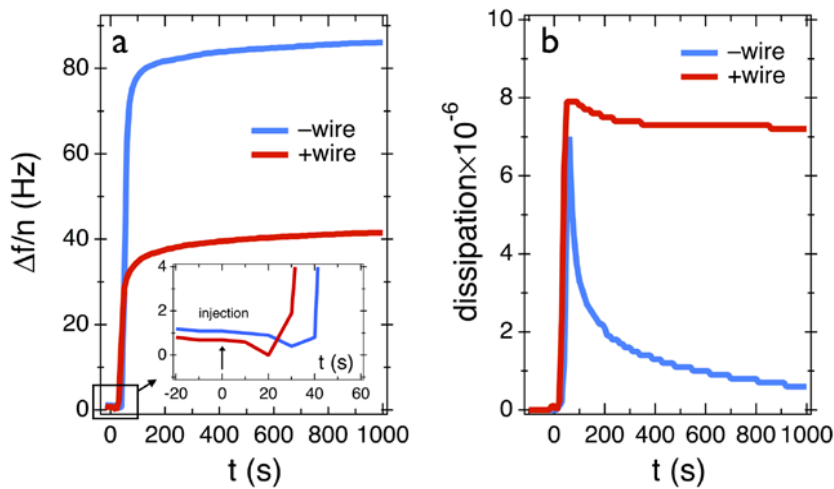

*Figure 6*: a) Frequency shift $\Delta f(t)$ for both negative and positive wires when exposed to a 0.48 wt. % SDS solution. The increase of the frequency shift after the injection of SDS is interpreted as a loss of materials from the surface. Inset: close-up view of the initial frequency shift showing a small but measurable negative undershoot. b) Time evolution of the dissipation $\Delta D$ associated with the behavior of Fig. 6a.

Effect of the concentration

The role of surfactant concentration on the swelling kinetics is also explored. Experiments are performed at SDS concentration $c_{SDS}$ = 0.048, 0.24, 0.48, 0.96, 4.8 and 30 wt. %, *i.e.* at 1/5, 1, 2, 4, 10 and 125 times the cmc, respectively. For concentrations above the cmc, the SDS molecules are in the state of spherical micelles (diameter 4 nm). Figs. 7a and 7b compare the swelling kinetics of wires treated with SDS solutions at $c_{SDS}$ = 0.24, 1 and 30 wt. % for positive and negative wires, respectively. The light grey lines in the figures represent the length increases of individual wires, whereas the colored lines are averages over 5 to 10 objects. Results at $c_{SDS}$ = 0.048 and 0.24 wt. % reveal no measurable length increase during the two hours of exposure to SDS, and no disassembly. For the $c$ = 0.48 and 4.8 wt. % sample, the data were comparable to those at 0.96 wt. %. For both positive and negative wires, $\Delta L_{Rel}(t)$ exhibits a behavior similar to those of Figs. 2 and 4: a rapid increase followed by a progressive leveling-off of the signal. For the 30 wt. % experiments, the length passes through a maximum before reaching its steady state. The transient profiles at this high concentration suggest that the swelling is arrested at a lower level because of the formation of extra structures between the polyelectrolytes, the nanoparticles and the large excess of SDS micelles present in the wire vicinity.

Fig. 7c shows the dependence of the relative length increase $\Delta L_{Rel}(c_{SDS})$ as a function of the SDS concentration for both types of wires. $\Delta L_{Rel}(c_{SDS})$ exhibits a discontinuity at low concentration, indicating the existence of a threshold. This threshold is estimated to be of the order of the cmc. Below, the wires remain in their pristine state. Above the threshold, $\Delta L_{Rel}(c_{SDS})$ passes through a broad maximum and decreases. Such a threshold was also observed with L*b*L multilayers treated with ionic surfactants.[16, 18] The existence of a threshold could be advantageously utilized to trigger the release of actives in biological environments.[10,

[11] Complementary studies using cationic surfactants (DTAB) and organic salt (NH$_4$Cl) in place of SDS reveal that the wires swell but do not disassemble. The data for the relative length increase are shown in Supplementaty Information (S6). These results indicate that the complexation between DTAB and PAA$_{2K}$ do not take place, and that competitive complexation depends on the pairs of oppositely charged species.[16, 18] Fig. 8 and Table I provide a summary of the $\Delta L_{Rel}$-results obtained in these different experimental conditions.

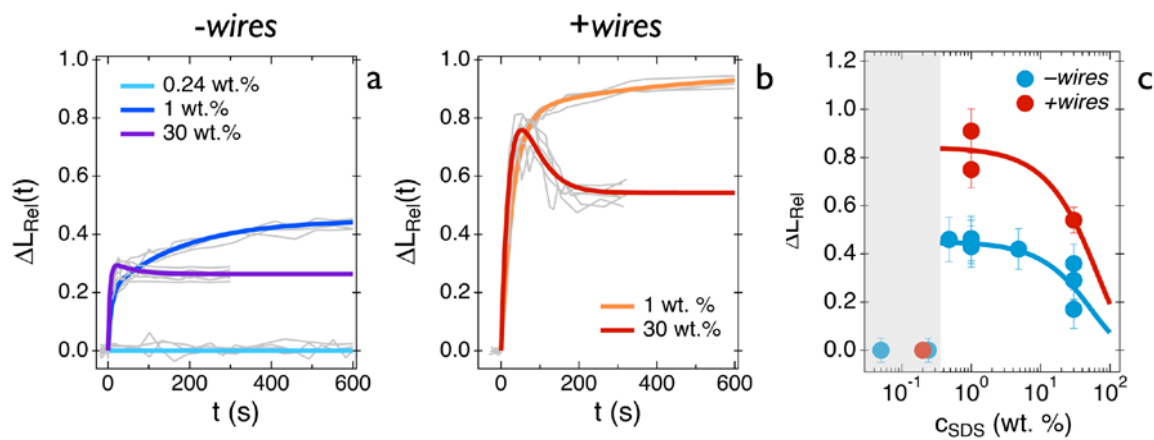

*Figure 7*: Relative length increase $\Delta L_{Rel}(t)$ of a) negative and b) positive wires observed at different SDS concentrations in the swelling regime. Data in grey are those of individual objects, and the colored lines represent the average. c) Stationary values of $\Delta L_{Rel}$ as a function of the SDS concentration for positive and negative wires. Both show a transition at $c_{SDS} \sim cmc$, i.e. 0.24 wt. %.

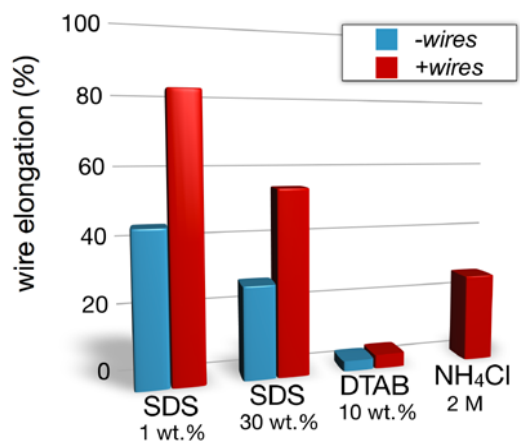

*Figure 8*: Comparison of the relative length increase $\Delta L_{Rel}$ in the swelling regime for the different experimental conditions studied here.

## Conclusion

The disassembly process described in this study bears strong similarities with that of L*b*L films[15, 17, 18] and capsules[16] examined in comparable conditions. We hence propose that the disassembly of iron oxide wires originate from the preferential complexation of the PDADMAC polycation with the SDS unimers or micelles. The displacement of the iron oxide particles by the surfactants suggests that polycation/surfactant complexes are more stable than those made with the polycation/polyanion pair.[10, 15, 16, 18, 26] Here we explore the kinetics of the processes involved in the disassembly (such as swelling and erosion) and provide a scenario for the overall mechanism. Our approach also allows for the real time monitoring of the erosion process at the microscopic level.

From the physico-chemical point of view, the impact of several parameters on the disassembly process was examined: the charges of the wires, the nature and the concentration of the ionic surfactants. To evaluate the role of electrostatic charges, positively and negatively charged wires with identical dimensions were prepared. Positive wires exhibit stronger swelling than negative ones due to their intrinsic different composition. However, the kinetics remains the same, indicating that charge repulsion between surfactants and wires does not interfere in the competitive binding exchange. By varying the SDS concentration from 1/5 to 125 times the cmc, a broad range of experimental conditions is also explored. Data show that there exists a threshold above which competitive binding occurs, and that this concentration is of the order of the cmc (0.24 wt. % for SDS). This result suggests that the $\gamma$-$Fe_2O_3$ particles forming the wire network are replaced by already preformed micelles, and not solely by unimers. The absence of swelling found with a cationic surfactant (DTAB) *in lieu* of SDS shows that the competitive binding is selective and depends on the pairs of oppositely species, a result already noticed in earlier reports.[16, 18]

Using phase-contrast optical microscopy and a controlled fluidic environment for the addition of surfactant, the interactions between the SDS and the wires reveal a two-stage process. The wires, either positive of negative swell very rapidly (some tens of seconds) and their structures soften. The large distortions of the particle/polymer matrix occurring during the swelling allow the further incorporation of micelles inside the structures. In the second stage, the erosion and detachment of 0.5 – 2 μm aggregates from the structure lead to the complete disassembly of the wires. If we take into account the QCM data, the description of the disassembly can be refined. With QCM, it is found that the frequency shift turns positive rapidly after the injection of SDS and exhibits a kinetic profile similar to that of the swelling. The agreement between the two behaviors indicates that even at the early stage, the swelling is accompanied by the release of nanoparticles in the surrounding, a result that could not be inferred from optical microscopy alone. Taken together, the data finally suggest that the erosion is not a bulk-type process, but is initiated at the surface, confirming the hypothesis of a surface-type swelling and erosion.[10]

# Acknowledgments

We thank Jérôme Fresnais for fruitful discussions. The Laboratoire Physico-chimie des Electrolytes, Colloïdes et Sciences Analytiques (UMR Université Pierre et Marie Curie-CNRS n° 7612) is acknowledged for providing us with the magnetic nanoparticles. ANR (Agence Nationale de la Recherche) and CGI (Commissariat à l'Investissement d'Avenir) are gratefully acknowledged for their financial support of this work through Labex SEAM (Science and Engineering for Advanced Materials and devices) ANR 11 LABX 086, ANR 11 IDEX 05 02.

## Supplementary Information

The Supporting Information includes sections on the characterization of iron oxide nanoparticle by light scattering and transmission electron microscopy (S1) and by vibrating sample magnetometry. Section S3 describes the process of the desalting transition and the fabrication of the nanostructured nanowires. S4 shows three movies of the disassembly of wires. The effect of pH on the nanowire stability is studied in S5, whereas S6 deals with the effect of the concentration, surfactant and salt on the swelling process. S7 shows images of wires on the QCM sensor silica surface by optical microscopy.

## References

1. Chapel, J. P.; Berret, J.-F. Versatile electrostatic assembly of nanoparticles and polyelectrolytes: Coating, clustering and layer-by-layer processes. *Curr. Opin. Colloid Interface Sci.* **2012,** *17* (2), 97-105.
2. Decher, G. Fuzzy Nanoassemblies: Toward Layered Polymeric Multicomposites. *Science* **1997,** *277*, 1232 - 1237.
3. Ostrander, J. W.; Mamedov, A. A.; Kotov, N. A. Two Modes of Linear Layer-by-Layer Growth of Nanoparticle-Polyelectrolyte Multilayers and Different Interactions in the Layer-by-Layer Deposition. *J. Am. Chem. Soc.* **2001,** *123*, 1101 - 1110.
4. Kataoka, K.; Harada, A.; Nagasaki, Y. Block Copolymer Micelles for Drug Delivery: Design, Characterisation and Biological Significance. *Adv. Drug. Del. Rev.* **2001,** *47*, 113 - 131.
5. Kataoka, K.; Togawa, H.; Harada, A.; Yasugi, K.; Matsumoto, T.; Katayose, S. Spontaneous formation of Polyion Complex Micelles with Narrow Distribution from Antisense Oligonucleotide and Cationic Copolymer in Physiological Saline. *Macromolecules* **1996,** *29*, 8556 - 8557.
6. Beija, M.; Marty, J. D.; Destarac, M. RAFT/MADIX polymers for the preparation of polymer/inorganic nanohybrids. *Prog. Polym. Sci.* **2011,** *36* (7), 845-886.
7. Wan, L.; Manickam, D. S.; Oupicky, D.; Mao, G. Z. DNA Release Dynamics from Reducible Polyplexes by Atomic Force Microscopy. *Langmuir* **2008,** *24* (21), 12474-12482.
8. Blacklock, J.; You, Y. Z.; Zhou, Q. H.; Mao, G. Z.; Oupicky, D. Gene delivery in vitro and in vivo from bioreducible multilayered polyelectrolyte films of plasmid DNA. *Biomaterials* **2009,** *30* (5), 939-950.
9. DeMuth, P. C.; Min, Y. J.; Huang, B.; Kramer, J. A.; Miller, A. D.; Barouch, D. H.; Hammond, P. T.; Irvine, D. J. Polymer multilayer tattooing for enhanced DNA vaccination. *Nature Mater.* **2013,** *12* (4), 367-376.
10. Lynn, D. M. Peeling back the layers: Controlled erosion and triggered disassembly of multilayered polyelectrolyte thin films. *Adv. Mater.* **2007,** *19* (23), 4118-4130.
11. Wohl, B. M.; Engbersen, J. F. J. Responsive layer-by-layer materials for drug delivery. *J. Controlled Release* **2012,** *158* (1), 2-14.
12. Westwood, M.; Gunning, A. P.; Parker, R. Temperature-Dependent Growth of Gelatin-Poly(galacturonic acid) Multilayer Films and Their Responsiveness to Temperature, pH, and NaCl. *Macromolecules* **2010,** *43* (24), 10582-10593.

13. Westwood, M.; Noel, T. R.; Parker, R. Environmental Responsiveness of Polygalacturonic Acid-Based Multilayers to Variation of pH. *Biomacromolecules* **2011,** *12* (2), 359-369.

14. Han, L. L.; Mao, Z. W.; Wuliyasu, H.; Wu, J. D.; Gong, X.; Yang, Y. G.; Gao, C. Y. Modulating the Structure and Properties of Poly(sodium 4-styrenesulfonate)/Poly(diallyldimethylammonium chloride) Multilayers with Concentrated Salt Solutions. *Langmuir* **2012,** *28* (1), 193-199.

15. Abdelkebir, K.; Gaudiere, F.; Morin-Grognet, S.; Coquerel, G.; Atmani, H.; Labat, B.; Ladam, G. Protein-Triggered Instant Disassembly of Biomimetic Layer-by-Layer Films. *Langmuir* **2011,** *27* (23), 14370-14379.

16. Kang, J.; Dahne, L. Strong Response of Multilayer Polyelectrolyte Films to Cationic Surfactants. *Langmuir* **2011,** *27* (8), 4627-4634.

17. Rahim, M. A.; Choi, W. S.; Lee, H. J.; Jeon, I. C. Ionic Surfactant-Triggered Renewal of the Structures and Properties of Polyelectrolyte Multilayer Films. *Langmuir* **2010,** *26* (7), 4680-4686.

18. Ramos, J. J. I.; Llarena, I.; Fernandez, L.; Moya, S. E.; Donath, E. Controlled Stripping of Polyelectrolyte Multilayers by Quaternary Ammonium Surfactants. *Macromol. Rapid Commun.* **2009,** *30* (20), 1756-1761.

19. Zahn, R.; Thomasson, E.; Guillaume-Gentil, O.; Voros, J.; Zambelli, T. Ion-induced cell sheet detachment from standard cell culture surfaces coated with polyelectrolytes. *Biomaterials* **2012,** *33* (12), 3421-3427.

20. Jing, J.; Szarpak-Jankowska, A.; Guillot, R.; Pignot-Paintrand, I.; Picart, C.; Auzély-Velty, R. Cyclodextrin/Paclitaxel Complex in Biodegradable Capsules for Breast Cancer Treatment. *Chem. Mater.* **2013,** *25* (19), 3867-3873.

21. Van Tassel, P. R. Polyelectrolyte adsorption and layer-by-layer assembly: Electrochemical control. *Curr. Opin. Colloid Interface Sci.* **2012,** *17* (2), 106-113.

22. Zelikin, A. N.; Quinn, J. F.; Caruso, F. Disulfide cross-linked polymer capsules: En route to biodeconstructible systems. *Biomacromolecules* **2006,** *7* (1), 27-30.

23. Marchenko, I.; Yashchenok, A.; Borodina, T.; Bukreeva, T.; Konrad, M.; Mohwald, H.; Skirtach, A. Controlled enzyme-catalyzed degradation of polymeric capsules templated on CaCO3: Influence of the number of LbL layers, conditions of degradation, and disassembly of multicompartments. *J. Controlled Release* **2012,** *162* (3), 599-605.

24. Inoue, H.; Sato, K.; Anzai, J. Disintegration of layer-by-layer assemblies composed of 2-iminobiotin-labeled poly(ethyleneimine) and avidin. *Biomacromolecules* **2005,** *6* (1), 27-29.

25. Sato, K.; Imoto, Y.; Sugama, J.; Seki, S.; Inoue, H.; Odagiri, T.; Hoshi, T.; Anzai, J. Sugar-induced disintegration of layer-by-layer assemblies composed of concanavalin a and glycogen. *Langmuir* **2005,** *21* (2), 797-799.

26. An, Y.; Bai, H.; Li, C.; Shi, G. Disassembly-driven colorimetric and fluorescent sensor for anionic surfactants in water based on a conjugated polyelectrolyte/dye complex. *Soft Matter* **2011,** *7* (15), 6873-6877.

27. Von Klitzing, R.; Wong, J. E.; Jaeger, W.; Steitz, R. Short range interactions in polyelectrolyte multilayers. *Curr. Opin. Colloid Interface Sci.* **2004,** *9* (1-2), 158-162.

28. Schlenoff, J. B.; Dubas, S. T. Mechanism of polyelectrolyte multilayer growth: Charge overcompensation and distribution. *Macromolecules* **2001,** *34* (3), 592-598.

29. Schonhoff, M. Self-assembled polyelectrolyte multilayers. *Curr. Opin. Colloid Interface Sci.* **2003,** *8* (1), 86-95.

30. Fresnais, J.; Berret, J.-F.; Frka-Petesic, B.; Sandre, O.; Perzynski, R. Electrostatic Co-Assembly of Iron Oxide Nanoparticles and Polymers: Towards the Generation of Highly Persistent Superparamagnetic Nanorods. *Adv. Mater.* **2008,** *20* (20), 3877-3881.

31. Yan, M.; Fresnais, J.; Berret, J.-F. Growth mechanism of nanostructured superparamagnetic rods obtained by electrostatic co-assembly. *Soft Matter* **2010,** *6* (9), 1997-2005.

32. Yan, M.; Fresnais, J.; Sekar, S.; Chapel, J. P.; Berret, J.-F. Magnetic Nanowires Generated via the Waterborne Desalting Transition Pathway. *ACS Appl. Mater. Interfaces* **2011,** *3* (4), 1049-1054.

33. Massart, R.; Dubois, E.; Cabuil, V.; Hasmonay, E. Preparation and properties of monodisperse magnetic fluids. *J. Magn. Magn. Mat.* **1995,** *149* (1-2), 1 - 5.

34. Berret, J.-F.; Sehgal, A.; Morvan, M.; Sandre, O.; Vacher, A.; Airiau, M. Stable oxide nanoparticle clusters obtained by complexation. *J. Colloid Interface Sci.* **2006,** *303* (1), 315-318.

35. Berret, J.-F. Stoichiometry of electrostatic complexes determined by light scattering. *Macromolecules* **2007,** *40* (12), 4260-4266.

36. Berret, J.-F.; Sandre, O.; Mauger, A. Size distribution of superparamagnetic particles determined by magnetic sedimentation. *Langmuir* **2007,** *23* (6), 2993-2999.

37. Fresnais, J.; Yan, M.; Courtois, J.; Bostelmann, T.; Bee, A.; Berret, J. F. Poly(acrylic acid)-coated iron oxide nanoparticles: Quantitative evaluation of the coating properties and applications for the removal of a pollutant dye. *J. Colloid Interface Sci.* **2013,** *395*, 24-30.

38. Evans, D. F.; Wennerström, K. *The Colloidal Domain*; Wiley-VCH: New York, 1999.

39. Vogt, B. D.; Soles, C. L.; Lee, H. J.; Lin, E. K.; Wu, W. L. Moisture absorption and absorption kinetics in polyelectrolyte films: Influence of film thickness. *Langmuir* **2004,** *20* (4), 1453-1458.

# TOC Image

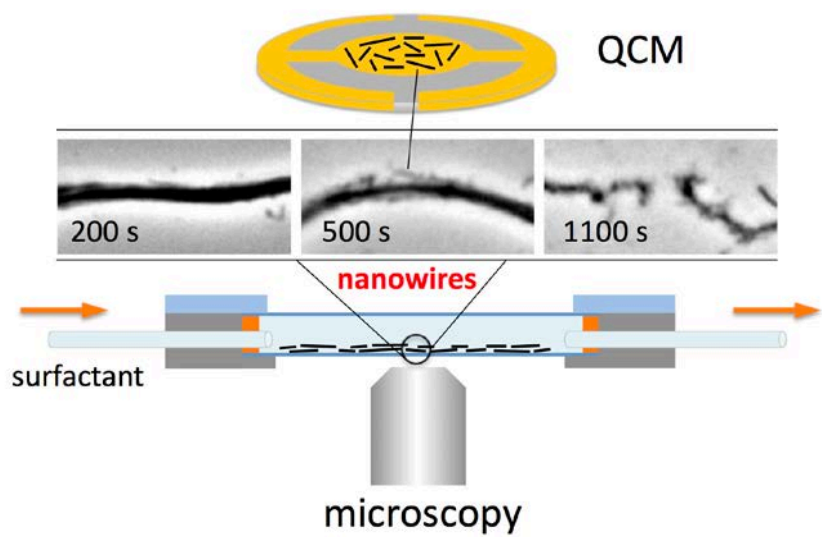

# Supporting Information

**Swelling and disassembly of electrostatic aggregates upon addition of surfactant**

N. Schonbeck[a]*, K. Kvale[a], T. Demarcy[a], J. Giermanska[b], J.-P. Chapel[b]* and J.-F. Berret[a]*

*Outline*

**S1 – Characterization of nanoparticle sizes and size distribution**
**S2 – Vibrating sample magnetometry**
**S3 – Desalting transition: towards the fabrication of nanostructured nanowires**
**S4 - Examples of movies of the disassembly of wires**
**S5 – Effect of pH on the nanowire stability**
**S6 – Effect of the concentration, surfactant and salt**
**S7 – Behavior of wires on the QCM sensor silica surface**

**S1 – Characterization of nanoparticle sizes and size distribution**

Electron beam microdiffraction experiments were performed on the iron oxide dispersion using a Jeol-100 CX transmission electron microscope at the SIARE facility of University Pierre et Marie Curie (Université Paris 6). The electron beam was focused on a selected area comprising a large number of nanoparticles, which diffraction pattern was recorded in the Fourier plane of the microscope. The reinforcement of the scattering along the rings is interpreted as arising from the diffraction by nanoparticles of different orientations with respect to the incoming beam. The wave-vector dependence of the intensity is shown in Fig. S1. The high crystallinity of the nanoparticles was evidenced by the observation of Bragg reflections that were indexed thanks to a calibration with gold and attributed to maghemite iron oxide ($\gamma$-$Fe_2O_3$). Table S1 provides the list of the Bragg wave-vectors and Miller indices for $\gamma$-$Fe_2O_3$. The positions of the Bragg peaks found experimentally for maghemite nanoparticles compare well with those predicted for this structure.

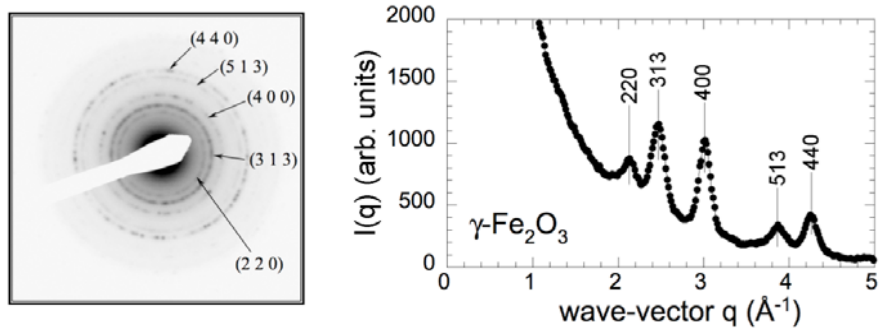

*Figure S1*: Microdiffraction spectrum obtained for iron oxide nanoparticles (left). Wave-vector dependence of the scattering intensity obtained for iron oxide nanoparticles (right).

| Theoretical values and proportions for γ-Fe$_2$O$_3$ | | | Experimental data |
|---|---|---|---|
| Miller indices for maghemite structure | $q_i$ (Å$^{-1}$) | Proportions | $q_i$ (Å$^{-1}$) |
| (2 2 0) | 2.1299 | 30 % | 2.13 |
| (3 1 3) | 2.5033 | 100 % | 2.47 |
| (4 0 0) | 3.0063 | 15 % | 3.01 |
| (4 2 6) | 3.6960 | 9 % | n.d |
| (5 1 3) | 3.9270 | 20 % | 3.86 |
| (4 4 0) | 4.2743 | 40 % | 4.26 |

*Table S1*: Theoretical Miller indices, Bragg wave-vectors and proportions for maghemite structure, and experimental wave-vectors for the peaks observed in Fig. S1. The theoretical data are from R.M. Cornell, U. chwertmann, 1996, The iron oxides structure, properties, reactions, occurrences and uses, V.C.H., Weinheim.

## S2 – Vibrating sample magnetometry (VSM)

Vibrating sample magnetometry (VSM) consisted in measuring the magnetization *versus* excitation M(H) for a solution at volume fraction $\phi$ from the signal induced in detection coils when the sample is moved periodically in an applied magnetic field (thanks to synchronous detection and with an appropriate calibration). Here we used the superparamagnetic nanoparticles having dominant populations at 8.3 nm. Fig. S2 shows the evolution of the macroscopic magnetization $M(H)$ normalized by its saturation value $M_S$ for the γ–Fe$_2$O$_3$ superparamagnetic nanoparticles. Here, $M_S = \phi m_S$, where $m_S$ is the specific magnetization of colloidal maghemite ($m_S = 3.5 \times 10^5$ A m$^{-1}$) and $\phi$ the volume fraction. The solid lines in Fig. S2 were obtained using the Langevin function for superparamagnetism convoluted with a log-normal distribution function of the particle size. The parameters of the distribution are the median diameter $D_0^{VSM} = 8.3$ nm and the polydispersity $s^{VSM} = 0.26$. These values are in relative good agreement with the ones obtained by TEM, albeit from a minor difference in diameter which could originate from defects located close to the surface and that would not contribute to magnetic properties.

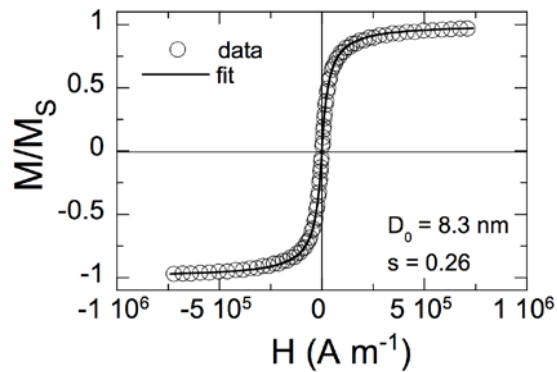

*Figure S2:* Magnetic field dependence of the macroscopic magnetization M(H) normalized by its saturation value $M_S$ for cationic maghemite dispersions. The solid curve was obtained using the Langevin function for superparamagnetism convoluted with a log-normal distribution function for the particle sizes, given with median

*diameters $D_0^{VSM}$ and polydispersity $s^{VSM}$.*

## S3 – Desalting transition: towards the fabrication of nanostructured nanowires

The protocols for mixing oppositely charged species described in this section were inspired by molecular biology and were developed for the *in vitro* reconstitutions of chromatin. The protocols applied here consisted first in the screening of the electrostatic interactions by bringing the dispersions of oppositely charged species to high salt concentration, and second in removing the salt progressively by *dialysis*. With this technique, the oppositely charged species were intimately mixed in solution but did not interact owing to the electrostatic screening. The dialysis strategy involved in a first step the preparation of separate 1 M salted solutions containing respectively the anionic particles and the cationic copolymers. The salt used in the present work was ammonium chloride ($NH_4Cl$). The two solutions were then mixed with each other. It was checked by light scattering that the stability of the dispersion was not changed by the presence of salt. In a second step, the ionic strength of the dispersions was progressively diminished. Dialysis was performed against deionized water using a Slide-a-Lyzer® cassette with molecular weight cut-off 10000 g mol$^{-1}$ (Fig. S3a). The time evolution of the ionic strength was monitored by the measurement of the electric conductivity of the bath. Dialysis was carried under two different conditions, with or without magnetic field. The cartoon in Fig. S3a represents the case where the magnetic field (0.3 T) was on. The nanowires were characterized by optical microscopy (inset of Fig. S3a) and by transmission electron microscopy (Fig. S3b).

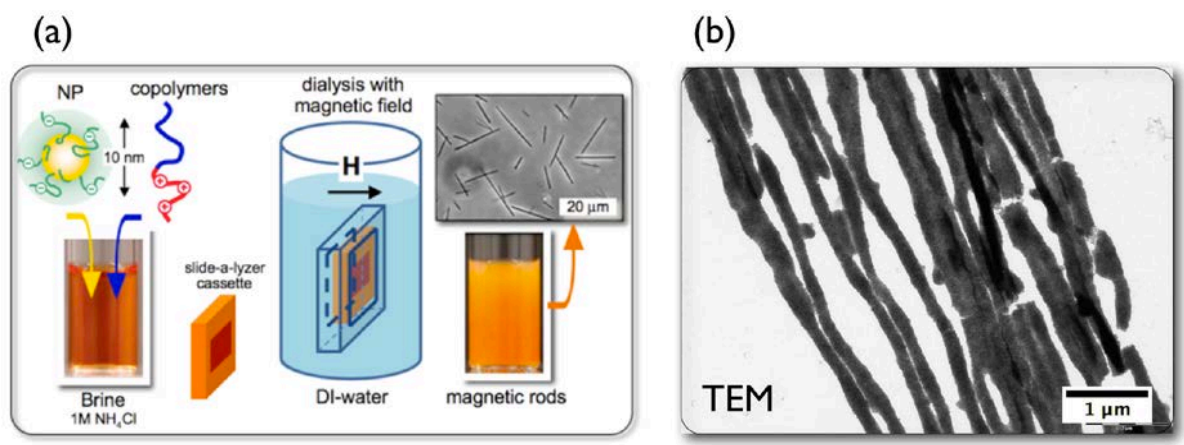

***Figure S3:*** *a) Schematic representation of the protocol that controls the nanoparticle co-assembly and wire formation. b) image of wires aggregated on a TEM grid.*

## S4 - Examples of movies of the disassembly of wires

List of the movies showing the disassembly of the wires, and posted as supporting material.

**movie#1 (figure2)**
**movie#2 (figure3)**
**movie#3 (figure5)**

## S5 – Effect of pH on the nanowire stability

Experiments were performed to test the integrity of the magnetic nanowires as a function of the *pH*. The *pH* of 15 μm nanowires dispersions was modified by addition of hydrochloric acid (pH 1.5, 3.4, 4.1) or sodium hydroxide (pH 9.1). The acidic conditions down to pH 4 reproduce some of the physico-chemical conditions encountered in late endosome/lysosome compartments. After 4 days, the wires were observed by phase-contrast microscopy and compared to the neutral pH conditions. Representative images are illustrated in Fig. S5.

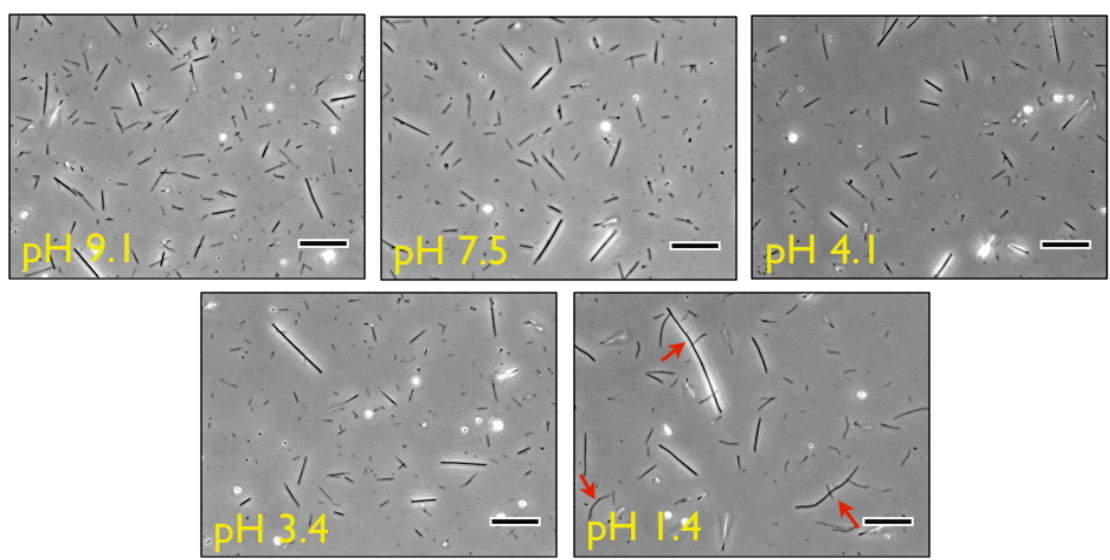

*Figure S5* : *Phase-contrast images of magnetic nanowires at different pH.*

Fig. S5 shows that from pH 9.1 down to pH 3.4 the wires remained intact, and comparable to those of neutral conditions (pH 7.5). At pH 1.4, the wires changed slightly but were still not degraded. Few wires exhibit kinks and bending that indicate a softening of their structure. It can thus be concluded that the degradation of the wires internalized by the fibroblasts was not due to the acidic pH conditions.

**S6 – Effect of the concentration, surfactant and salt**
The swelling was further studied using different types of injectant, such as cationic surfactants and organic salt. Fig. S6 compares the swelling behavior of +*wires* subjected to anionic (SDS, 0.96 wt. %) and cationic (DTAB, 10 wt.%) solutions. It is found that the cationic surfactants have almost no effect on the wires. The lengths increased by 2-3%, and results were similar for +*wires* and −*wires*. Wires subjected to the organic salt ($NH_4Cl$, 2M) exhibited significant swelling. With cationic surfactant and with organic salt however, the initial growth velocity (on the order of 0.05 μm s$^{-1}$) remains 10 to 100 times smaller than for wires subjected to SDS. Additionally, although the wires in contact with organic salt become flexible, they did not disassemble.

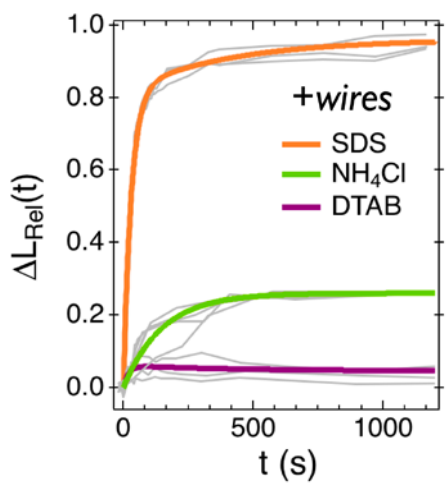

***Figure S6***: *Relative length increase $\Delta L_{Rel}(t)$ of positive wires submitted to solutions containing anionic (SDS, 0.96 wt. %), to cationic (DTAB, 10 wt.%) and to 2 M NH$_4$Cl solutions. Data in grey are those of individual objects, and the colored lines represent the average.*

With DTAB, the wires remained close to their original state, with a swelling reduced to less than 3% and no increase in flexibility. Results were identical for $+wires$ and $-wires$, which indicates that the complexation, here between the surfactant and the PAA$_{2K}$-coated particles did not take place, and that competitive complexation depends on the pairs of oppositely charged species. Beyond the neat difference observed between SDS and DTAB molecules onto macroscopic wires, it should be noted that the same tendency was observed on a model layer by layer system made from alternating 7 nm cerium oxide nanoparticles and PDADMAC layer (data not shown) where adsorption and swelling occurred upon injection of SDS whereas nothing ever happened in case of DTAB.

## S7− Behavior of wires on the QCM sensor silica surface

The silica surface of the QCM sensor was washed with ethanol, dried with N$_2$ and treated by 20 min UV. For already used sensors, the surface was washed in an ultrasonic bath containing a 4 M NaCl solution for one hour and rinsed with water. In a second step, the sensor was dipped in a KOH/ethanol mixtures for 30min, rinsed with ethanol, water and finished with 20 min UV. Dispersions of positive (Z = 0.3) or negative (Z = 7) wires were diluted down to c = 0.01 wt. %, and a drop of this solution was deposited on the sensor and left to dry in the air. Fig. S7 displays images of positive and negative wires before the SDS treatment and 5 hours after (objective 40X). Here, most of the wires disappeared but some partially eroded and barely seen remain, a result that could be attributed to the preparation of the seeded sensor.

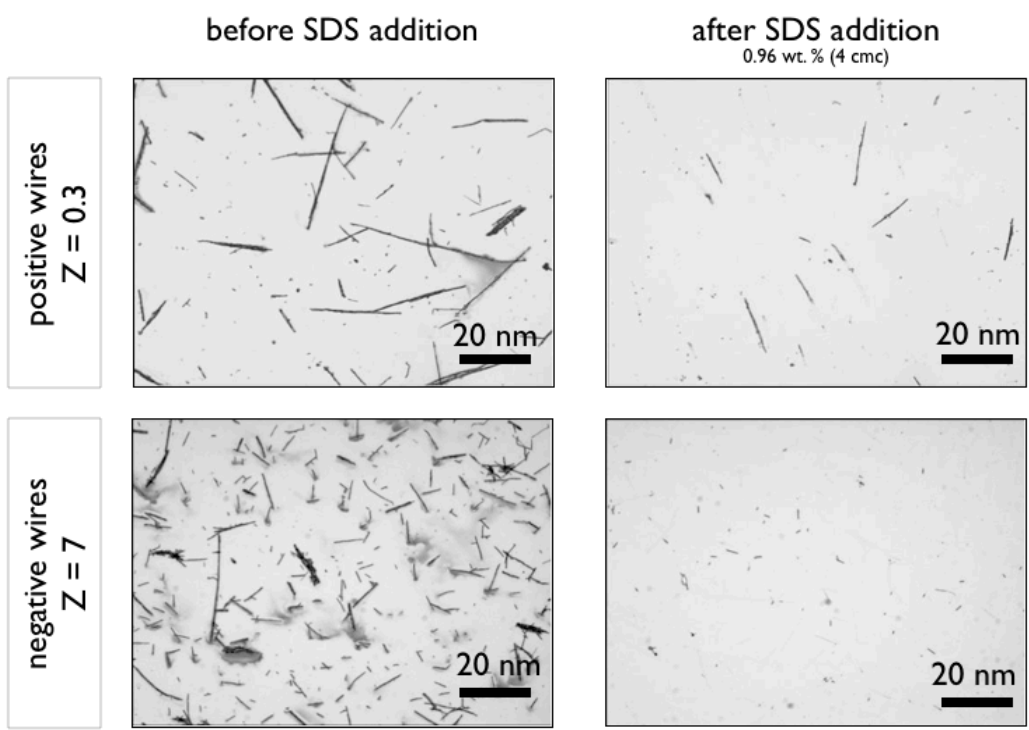

*Figure S7: Optical microcopy images of positive and negative wires deposited on the quartz crystal surface before and after running the QCM experiment.*